\documentstyle[aps,prc,epsfig]{revtex}

\def\lp {\left( }
\def\rp {\right) }
\def\lb {\left[ }
\def\rb {\right] }
\def\lc {\left\{ }
\def\rc {\right\} }

\def\nn {\nonumber}

\def\beq{\begin{equation}}
\def\eeq{\end{equation}}
\def\bea{\begin{eqnarray}}
\def\eea{\end{eqnarray}}
\def\ni{\noindent}

\def\d {\partial }

\def\cd {\!\cdot\!}

\def\rar {\rightarrow}

\def\Db {\bar{D}}

\def\a{\alpha}

\def\e{\epsilon}

\def\G {\Gamma}

\def\L {\Lambda}
\def\m{\mu}
\def\n{\nu}

\def\p{\pi}
\def\P{\Pi}
\def\r{\rho}

\def\cD {{\cal D}}

\def\cL {{\cal L}}

\def\bro {\mbox{\boldmath $\rho$}}
\def\bT {\mbox{\boldmath $T$}}

\def\btau {\mbox{\boldmath $\tau$}}

\def\btau {\mbox{\boldmath $\tau$}}

\def\maior{\smash{\mathop{>}\limits_{\raise4pt\hbox{$\scriptstyle \sim$}}}}
\def\menor{\smash{\mathop{<}\limits_{\raise4pt\hbox{$\scriptstyle \sim$}}}}

\begin{document}

\title{ \bf {Meson loops and the $g_{D^\ast D\pi}$ coupling}}
\author{F.O. Dur\~aes$^{1,2}$\thanks{e-mail: fduraes@if.usp.br}, \, F.S.
Navarra$^{1}$\thanks{e-mail: navarra@if.usp.br}, \, M. Nielsen$^{1}$
\thanks{e-mail: mnielsen@if.usp.br} and M.R. Robilotta$^{1}$\thanks{e-mail: mane@if.usp.br}  
\\[0.1cm]
{\it $^1$Instituto de F\'{\i}sica, Universidade de S\~{a}o Paulo,}
\\[0.1cm]
{\it C.P. 66318, 05389-970 S\~{a}o Paulo, SP, Brazil} \\[0.1cm]
{\it $^2$ Dep. de F\'{\i}sica, Faculdade de Ci\^encias Biol\'ogicas, 
Exatas e Experimentais,} \\[0.1cm]
{\it Universidade Presbiteriana Mackenzie, C.P. 01302-907 S\~{a}o Paulo, 
Brazil}}
\maketitle
\vspace{1cm}
\begin{abstract}
The $D^*D\pi$ form factor  is evaluated at low and moderate $Q^2$ in a 
hadronic loop model, for off-shell  $D$ mesons. The results contain arbitrary 
constants, which are fixed so as to match previous QCD sum rule calculations 
valid at higher $Q^2$. The form factor obtained in this way was  used to extract 
the coupling constant, which is in very good agreement with the experimental value.

\end{abstract}

\vspace{1cm}

\section{Introduction}

The measurement of the  $D^*D\pi$ coupling, made  by the CLEO collaboration \cite{cleo}, 
yelding the result $g_{D^*D\pi}= 17.9 \pm 0.3 \pm 1.9 $, created embarrassement in the community of QCD sum rules (QCDSR) which, by means of various schemes, had predicted much smaller values. 
Several different approaches of QCDSR were employed: two point function combined with soft pion techniques \cite{col,ele}, light cone sum rules(LCSR) \cite{bel,col2}, light cone sum 
rules including perturbative corrections \cite{kho}, sum rules in a external 
field \cite{gro}, double momentum sum rules \cite{dn} and  double Borel sum rules \cite{ddpi}. 
The LCSR prediction made in ref.~\cite{bel} became even smaller after the radiative corrections have 
been included~\cite{kho}. 
The upper limit of these predictions was $g_{D^*D\pi}=13.5$ \cite{kho} and it turned out to be 30\% lower than the central value of the CLEO measurement.

Although the QCD sum rule approach certainly suffers from large uncertainties, in several other cases good agreements 
with experiment were obtained. 
Therefore, we cannot simply be skeptical about the whole sum rule approach. 
The $g_{D^*D\pi}$ coupling constant  does not seem, a priori, to be particularly exotic and other theoretical 
approaches did not produce such large discrepancies for this quantity. 
A careful discussion meant to reduce the uncertainties presented by quark models, performed in the framework of Dirac equation~\cite{dirac}, and prior to the experimental measurement, has led to the result $g_{D^\ast D \pi} \simeq 18$. 
It should be stressed that this result has been obtained in the heavy quark limit.  
The recent (quenched) lattice QCD calculation has produced
$g_{D^\ast D \pi} = 18.8 \pm 2.3^{+1.1}_{-2.0}$~\cite{herdoiza}. 
It is therefore important to understand the specific difficulty which the standard sum rule approach seems 
to encounter in this case.

After the appearance of experimental data, three works \cite{damir,kimlee,kim} tried to 
reconcile the LCSR estimates with the measured figure.
In \cite{damir}, it was noted that the inclusion of an explicit radial excitation contribution to the hadronic 
side of the LCSR (often referred to as the left hand side of the 
sum rule)  could  significantly improve the value of $g_{D^\ast D \pi}$ and, at the same time, the stability of the sum rule with respect to the Borel parameter $M^2$.

In the standard QCDSR approach of \cite{ddpi}, a modification in the continuum contribution 
(such as the explicit inclusion of a radial exictation) does not seem to be neither necessary nor promising, 
because there the Borel suppression is much more effective.
In ref.~\cite{ddpi} the $g_{D^*D\pi}(Q^2)$ form factor was estimated as  
a function of the off-shell pion momentum $Q^2$. Since the sum rule obtained from the three-point function adopted is not valid at $Q^2=0$, in order to determine
the $D^*D\pi$ coupling, it was necessary to extrapolate the $Q^2$ behaviour of the form factor. 
Of course there are large uncertainties in this procedure, and, the value of $g_{D^*D\pi}=5.7\pm 0.4$ obtained 
was much smaller than the experimental result.
In a subsequent calculation of the $DD\rho$ vertex \cite{ddrho}, the $DD\rho$ form factor was calculated for both 
$D$ and $\rho$
off-shell mesons and the QCDSR results were parametrized by analytical forms such that the respective extrapolations to the $D$ and $\rho$ poles provided consistent values for the $g_{DD\rho}$ coupling constant. 
This method of double parametrization plus matching at the on-shell point was then employed in \cite{revisit} 
to recalculate the $D^*D\pi$ coupling and led to the value 
\beq
g_{D^*D\pi}=14.0 \pm 1.5\;.
\label{gd}
\eeq

While this number is much closer to the experimental value, there is still a discrepancy. 
Moreover, the procedure of fitting the QCDSR points in the deep euclidean region and extrapolating them to 
the time-like region still contains uncertainties, such as, for example {\it the analytical form chosen for the parametrizations}, i.e., monopole, exponential or gaussian.

In the present work we return to this question and employ hadronic loops, calculated by means of effective field theories (EFT), 
in order to produce a better parametrization for $D^* D \pi$ results derived by means of QCDSR.
Purely hadronic calculations are independent from QCDSR and involve the choice of an effective Lagrangian, 
including the possible requirements of chiral symmetry and/or $SU(4)$. 
However, beyond tree level, one has to deal with the problems and uncertainties associated with renormalization.
As we discuss in the sequence, a suitable combination of  EFT and QCDSR results allows the elimination of 
undesired indeterminacies of both approaches, improving significantly their predictive powers.
Effective interactions are discussed in section II, whereas results and conclusions are presented in sects. IV and V.
As an straightforward exercise we also make a prediction for $g_{B^*B\pi}$. 

\section{Effective Dynamics}

The full $D^* D \pi$ vertex function is shown in Fig. 1.
Leading contributions to this vertex come from both the tree interaction and the three classes of diagrams depicted in 
Figs. 2, 3 and 4. 
Meson loops are a necessary consequence of quantum field theory and do contribute to several hadronic observables. 
In practice, due to problems associated with infinities, renormalization becomes unavoidable in the 
evaluation of loop corrections to observables.
Nowadays, this kind of procedure is rather well established at the hadronic level,
in processes such as pion-pion and pion-nucleon scatterings.
In the case of $D$ mesons, on the other hand, the theory is much less developed and hence we resort to an alternative.
The basic idea is to isolate the unknown loop parameters into some basic constants, in such a way that they can 
be determined by matching the results of  loop and QCDSR results. 

Before proceding, some remarks are in order.
We first note that some diagrams, such as, for instance,  that in Fig. 2a , contain internal vertices 
involving the $D^* D \pi$ coupling. 
This suggests that the calculation is ``cyclic'', since one needs to use the $D^* D \pi$  form factor in order to 
calculate the $D^* D \pi$  form factor. 
Actually, there are differences between the internal particles and the external ones. 
The former are always virtual, whereas the latter may be either real or put on mass shell in the extraction of the 
coupling constant. 
In the framework of perturbation theory, at leading order, internal particles are treated as elementary, without structure.
They are assumed to be point-like and the evaluation of leading terms does not require the use of internal form factors. 
Consistently, one must use bare coupling constants for these interactions.

There are heavy mesons circulating in the loops shown in Figs. 2-4 and one might be tempted to argue 
that other states should also be included. 
We do have, for example, fermion-antifermion components such as $\bar{N}N$ or $\bar{\Lambda}_c\Lambda_c$ 
in the loops.
An incoming positive pion can split into a $p$ plus a $\bar{n}$, and so on.
However, in a different context \cite{speth},  it has been shown that this kind of splitting is suppressed with respect 
to the pion $\rightarrow$ meson-meson splitting, by one order of magnitude. 
The neglect of this kind of contribution seems therefore justified.
The same holds for the possibility of strangeness circulating in the loop, associated with  virtual states 
such as $D_s$, 
$D^*_s$, $K$ and $K^*$. 
Using only $\pi$'s, $\rho$'s, $D$'s and $D^*$'s we cover the  low and  high $Q^2$ regions of the form factor. 
Thus it is enough to our purposes to work with a simple effective theory, 
involving only  $\pi$'s, $\rho$'s, $D$'s and $D^*$'s that, as has been discussed elsewhere, \cite{linko},  has proven to be phenomenologically successful.
The same happens in this work.

The diagrams considered in this calculation have been divided into three classes. 
The first one, represented in Fig. 2, involves only triangular loops.
The processes in Fig. 3, on the other hand, contain bubble-type loops and four-leg vertices associated with the gauge
structure of $\rho$ interactions.
Finally, diagrams given in Fig. 4 involve bubble loops as well as single-particle propagators.
This last feature might suggest that these interactions should be considered as mass corrections.
However, the nature of the effective interactions described below is such that, in some terms, the poles of the 
single-particle states are cancelled. 
These terms do correspond to proper three-point functions and are kept in the evaluation of the form factor.
Pole cancellations are indicated by crosses in Fig. 4.

We adopt an  effective Lagrangian constrained by $SU(2)$ flavor and chiral symmetries, as well as gauge invariance.  
The $\p D D^*$ interaction is given by\cite{linko,mamu98,osl,mane1}

\beq
\cL_{\p D D^*} = i \;\hat{g}_{\p D D^*}\lb \Db \tau_a D_\m^* - \Db_\m^* \tau_a D \rb \d^\m \phi_a\;.
\label{1}
\eeq

\ni
where $\tau_a$ are the Pauli matrices,  
$\phi_a$ denotes the pion isospin triplet, 
while $D\equiv (D^0,D^+)$ and $D^*\equiv (D^{*0},D^{*+})$  represent the pseudoscalar and vector 
charm meson doublets, respectively. The {\em hat} on top of the coupling constant indicates its bare nature.

The $\r$ couplings are assumed to be universal and are implemented by covariant derivatives of the form

\beq
\cD^\m = \d^\m - i \hat{g}_\r \;\bT\cd \bro^\m\;,
\label{2}
\eeq

\ni
where $\hat{g}_\r$ is the universal coupling constant and $\bT$ is the isospin matrix suited to 
the field $\cD^\m$ it is acting upon. In this work we need

\bea
&& \cD^\m D = \lb \d^\m - i \hat{g}_\r \;\frac{\btau}{2}\cd \bro^\m \rb D\;,
\label{3}\\[2mm]
&& \cD^\m \Db = \Db \lb \d^\m + i \hat{g}_\r \;\frac{\btau}{2}\cd \bro^\m \rb\;,
\label{4}\\[2mm]
&& \cD^\m \phi_a = \lb \d^\m +  \hat{g}_\r \;\e_{abc}\r_b^\m \rb \phi_c \;.
\label{5}
\eea

Using this prescription in eq. (\ref{1}), we obtain

\beq
\cL_{\r \p D D^*} = i \;\hat{g}_\r \; \hat{g}_{\p D D^*}\lb \Db \tau_b D_\m^* - \Db_\m^* \tau_b D 
\rb \e_{bca}\;\r_{c}^\m \phi_a\;.
\label{6}
\eeq

For the other couplings, we depart from the free Lagrangians and have

\bea
\cL_{\p\p} &=& \frac{1}{2}\; \d_\m \phi_a \;\d^\m \phi_a 
\rar  \cL_{\r \p\p}= \hat{g}_\r\; \e_{abc}\;\phi_a \d_\m \phi_b \;\r_c^\m \;,
\label{7}\\[2mm]
\cL_{\Db D} &=& \d_\m \Db \;\d^\m D 
\rar \cL_{\r DD}= \frac{i}{2}\;\hat{g}_\r \lb \Db\; \tau_c \; \d_\m D - \d_\m \Db \; \tau_c \; 
D \rb \r_c^\m \;,
\label{8}\\[2mm]
\cL_{\Db^* D^*} &=& - \frac{1}{2}\lb \d_\m\Db_\n^* -\d_\n \Db_\m^* \rb \lb \d^\m 
D^{*\n} -\d^\n D^{*\m} \rb \rar
\nn\\[2mm] 
\cL_{\r \Db^* D^*} &=& - \frac{i}{2}\;\hat{g}_\r \,\, [ \,\, \Db^{*\n}\tau_c \lp \d_\m D_\n^* 
-\d_\n D_\m^*\rp
- \lp \d_\m \Db_\n^* -\d_\n \Db_\m^*\rp \tau_c D^{*\n} \,\, ] \,\,  \r_c^\m \;.
\label{9}
\eea

With these Lagrangians we can write and evaluate the contributions of Figs. 2-4 to 
the total vertex function.

\section{Results}

The $\p^a(q)\;D(p)\; D_\a^*(p')$ vertex function $\G_\a^a(p^2)$ for an off-shell $D$ is written as 

\bea
\G^a_\a (p^2) = - \; \tau^a \; q_\a G(p^2) \;,
\label{3.1}
\eea

\ni
where $G(p^2)$ is a form factor, such that the physical coupling constant 
is $g_{\p D D^*}=G( m_D^2)$.
We consider two kinds of  loop corrections  to this vertex, containing pion and rho intermediate states,
denoted respectively by $F_\p(p^2)$ and $F_\r(p^2)$.
The perturbative evaluation of these functions gives rise to divergent integrals and $G(p^2)$ can be determined 
only up to yet unknown renormalization constants. 

The use of standard loop integration techniques, such as dimensional regularization and $\overline{MS}$ 
subtraction of divergences, for all diagrams, allows one to write the form factor as 

\bea
G(p^2) = K+ C_\p \;F_\p( p^2) + C_\r \;F_\r (p^2) \;,
\label{gfinal}
\eea

\ni
where $K$, $C_\p$ and $C_\r$ are constants.
These constants incorporate the bare couplings $g_0$ and $\hat{g}_\r$,
the usual parameters associated with renormalization and, in this work,
are determined by comparing the general structure of $G(p^2)$ with the results from QCD sum 
rules.

Keeping only the terms which depend on $p^2$, the explicit evaluation of the diagrams given 
in Figs. 2-4  yields:

\bea
F_\p &=&  \frac{1}{ (4\p)^2} \lc \lb (m_{D^*}^2-m_D^2)
-\;\frac{1}{4 m_{D^*}^2}\;(m_{D^*}^2-m_D^2+\m^2)^2 \rb \;\P_{\p D D^*}^{(001)}\right.
\nn\\[2mm]
&-& \left. 
\frac{1}{4 m_{D^*}^2}\;\lb  (m_{D^*}^2 + m_D^2 - \m^2)\;\P_{\p D^*}^{(01)}
+ 6 \;(m_{D^*}^2-m_D^2+\m^2)\;\P_{\p D^*}^{(00)} \rb \rc\;,
\label{fpi}
\eea

\bea
F_\r &=& \frac{1}{(4\p)^2} \; \lc - \lp 2 m_{D^*}^2 - 2 m_D^2 - 2 \m^2 + m_\r^2 \rp 
\lb \P_{\r\p D}^{(000)} + \P_{\r\p D}^{(010)} + \P_{\r\p D}^{(001)}\rb \right.
\nn\\[2mm]&+& \left. \lp m_{D^*}^2 - m_D^2 + \m^2  \rp 
\lb 2 \P_{\r\p D^*}^{(000)}+ \P_{\r\p D^*}^{(001)}\rb
+ 2 \lp m_{D^*}^2 - m_D^2 \rp  \lb \P_{\r\p D^*}^{(010)}\rb   \right.
\nn\\[2mm]
&-& \left. \frac{1}{2}\lp m_{D^*}^2 +  m_D^2 - \m^2  \rp \lb \P_{\r D D^*}^{(000)}\rb
-\frac{1}{8} \lp m_{D^*}^2 + 3 m_D^2 - \m^2  \rp \lb \P_{\r D D^*}^{(010)} \rb \right.
\nn\\[2mm]
&-&  \left. \frac{1}{8} \lp - m_{D^*}^2 + 5 m_D^2 -5 \m^2 \rp \lb 
\P_{\r D D^*}^{(001)}\rb
+  \frac{3}{8} \P_{\r D}^{(00)} \rc \;.
\label{fro}
\eea

The functions $\P_{xyz}$ and $\P_{xy}$ entering these results are Feynman integrals, with lower labels indicating 
the intermediate propagating states.
Upper indices represent the Lorentz tensor structure, which is realized in terms of the external variables $q$, $p$ and $p'$,
and defined by the relationships 

\bea
%
\P_{xyz}^{(klm)} &=& - \int_0^1d a\;a \int_0^1 d b \; [-(1\!-\!a)]^k \; [- a (1\!-\!b) ]^l  \; [-ab]^m\;
\lb 1 /D_{xyz}\rb  \;,
\nn\\[2mm]
D_{xyz} &=& (1\!-\!a) \;m_x^2 + a(1\!-\!b)\; m_y^2 + ab\;m_z^2 
\nn\\[2mm]
&-& a(1\!-\!a)(1\!-\!b)\;(p_x\!-\!p_y)^2 - a(1\!-\!a)b\;(p_x\!-\!p_z)^2 - a^2 b (1\!-\!b)\;(p_y\!-\!p_z)^2 \;,
\label{triangle}
\eea

\ni
and

\bea
%
\P_{xy}^{(kl)} &=& - \int_0^1d a\; [-(1\!-\!a)]^k \; [- a ]^l \; 
\ln \lb D_{xy}/ \bar{D}_{xy}\rb  \;,
\nn\\[2mm]
D_{xy} &=& (1\!-\!a) \;m_x^2 + a\; m_y^2 - a(1\!-\!a)\;(p_x\!-\!p_y)^2\;,
\nn\\[2mm]
\bar{D}_{xy} &=& (1\!-\!a) \;m_x^2 + a\; m_y^2\;.
\label{bubble}
\eea

Using (\ref{fpi}) and (\ref{fro}) into (\ref{gfinal}), we obtain the form factor as a function of $p^2$, the $D$ four-momentum squared. 
At this stage, it still contains three unknown parameters, which are determined by adjusting the function $G(p^2)$ to 
the QCD sum rule points taken from ref. \cite{revisit}.
Those results are displayed in Fig. 5, where $P^2 \equiv -p^2$, together with our best  fit ($\chi^2\sim 10^{-3}$) represented by the solid line.
Computing the value of $G(p^2)$ at $p^2=m_D^2$, we arrive at the following value for the coupling constant:

\beq
g_ {D^* D \pi} = 17.5 \pm 1.5 \;,
\label{Dresult}
\eeq
in very good  agreement with experiment.
The errors quoted come from the QCDSR points, which contain a typical error of $\simeq$ 10 \%.
In the same figure we also show the results of the fits of the QCDRS points with 
two mixed monople-dipole structures with three free parameters, namely 

\bea
G^{FI}(p^2) &=& C \lb \frac{\L_1^2-m_D^2}{\L_1^2-p^2} + \lp \frac{\L_2^2-m_D^2}
{\L_2^2- p^2}\rp^2 \rb \;,
\label{GFI}\\[2mm]
G^{FII} (p^2) &=& C_1 \, \frac{\L^2-m_D^2}{\L^2-p^2} + C_2\,\lp \frac{\L^2-m_D^2}
{\L^2- p^2}\rp^2  \;,
\label{GFII}
\eea

\ni
which yield $\chi^2_I\sim 10^{-3}$ (dashed line) and $\chi^2_{II}\sim 10^{-2}$ (dash-dotted line), respectively.
Inspecting Fig. 5 one learns that these alternative structures, reasonable as they  are, diverge significantly
from the loop calculation  in the region where the $D$ is not too off-shell, stressing
the importance of a proper hadronic treatment of the form factor in that region.

In this work we also consider the $B^* B \pi$ vertex with an off-shell $B$, which can be obtained by a straightforward
replacement of the charmed particles with those containing the quark $b$ in the results described above.
Using the same Lagrangians and fitting procedure with results from ref. \cite{revisit}, we obtain the curve shown in Fig. 6. 
The resulting coupling constant is:

\beq
g_ {B^* B \pi} = 44.7 \pm 1.0 \;,
\label{Bresult}\eeq

\ni
in  good agreement with lattice estimates.

As far as practical applications are concerned, our numerical results for the  form factors 
$G_{D^* D \pi}(p^2)$ and $G_{B^* B \pi}(p^2)$, in the whole range $m^2_{D(B)} \leq p^2 < -5$ GeV,
are very well described by the mixed monopole-dipole structure given by eq.(\ref{GFI})
with the parameters 

\bea
&& G_{D^* D \pi}(p^2) \rar C = 8.7\,\,;\,\, \Lambda_1=5.1 \,GeV 
\,\,;\,\,\Lambda_2=2.9\,GeV\;,
\nn\\[2mm]
&& G_{B^* B \pi}(p^2) \rar C = 22.4\,\,;\,\, \Lambda_1=7.8\,GeV \,\,;\,\,
\Lambda_2=7.1\,GeV\;.
\eea

Our good results allow us to believe that the use of meson loops can significantly reduce the uncertainty in the 
extrapolation of form factors, computed in the space-like region by means of QCDSR, to the time-like region,
with the corresponding increase in the reliability of predictions for coupling constants.  
It is worth stressing that, apart from the approximations described in the introduction, 
our procedure has no new source of errors.

\section{Conclusions}

We have developed a new method of improving QCDSR calculations of hadronic form factors, which 
consists in matching  QCDSR results, valid mainly in the deep euclidean region, to meson loop calculations,
valid when the $D$ is not too off-shell.
This matching is well justified from the physical point of view,  since in the intermediate and  large $Q^2$ regions
the relevant degrees of freedom are the quarks and gluons,  with non-perturbative corrections taken into 
account through the QCD condensates.  
The opposite happens  for low values of $Q^2$, where sum rules calculations become non-reliable due to the lack 
of a large mass scale. 
At this point, the meson exchange dynamics  becomes the most reliable tool,
but it depends on unknown constants associated with the renormalization of the mesonic vertices.  
Although the exact frontier between  meson dynamics and  QCDSR cannot be precisely known,
the success of the method in the instances considered here supports the view that the matching
may become useful in increasing the predictive power of both procedures.
This encourages us to reconsider our previous form factor studies.

\vspace{0.5cm}

\underline{Acknowledgements}: This work has been supported by CNPq and  
FAPESP under contract number 1999/12987-5.

   


\begin{figure} [h]
\centerline{\epsfig{figure=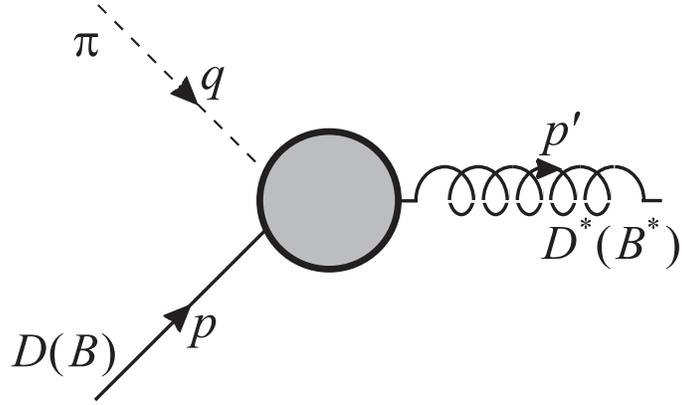,width=9.cm}}
\caption{The full $D^* D \pi$ ($B^* B \pi$) form factor.}
\label{fig1}
\end{figure}

\begin{figure} [h]
\centerline{\epsfig{figure=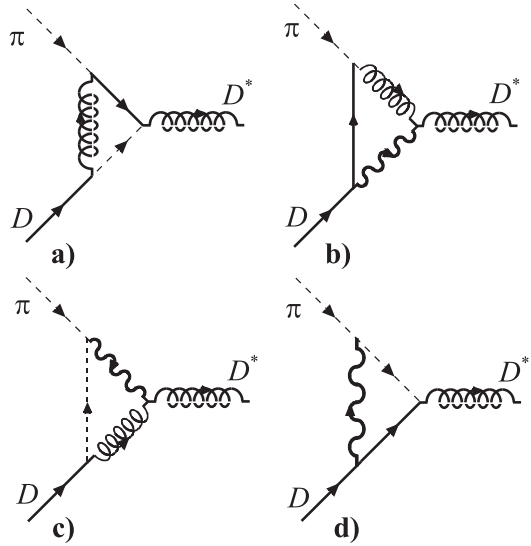,width=7.cm}}
\caption{Meson loop contributions to the $D^* D \pi$ form factor: 
``triangle'' diagrams.}
\label{fig2}
\end{figure}

\begin{figure} [h]
\centerline{\epsfig{figure=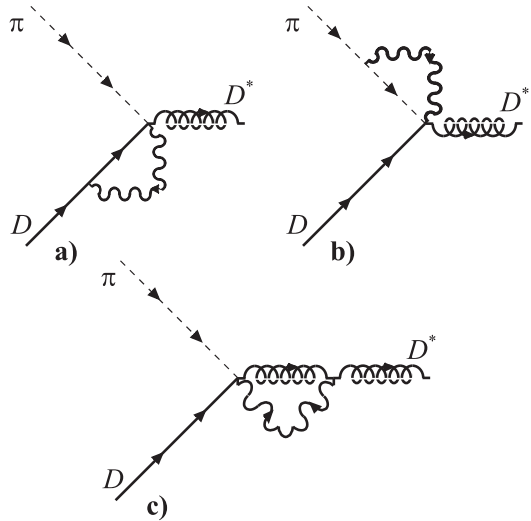,width=7.cm}}
\caption{Meson loop contributions to the  $D^* D \pi$ form factor: 
``quartic couplings''.}
\label{fig4}
\end{figure}

\begin{figure} [h]
\centerline{\epsfig{figure=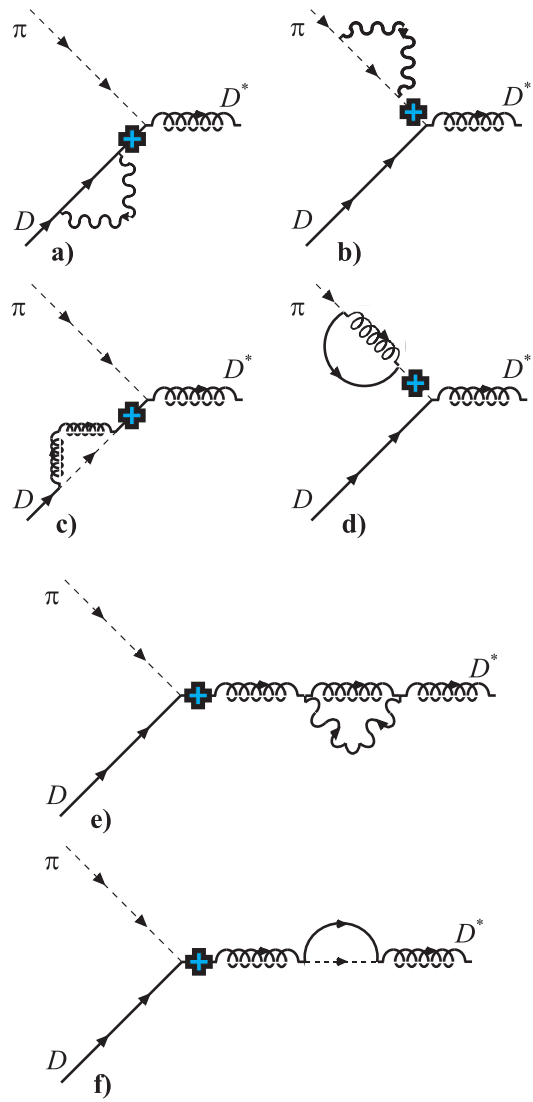,width=7.cm}}
\caption{Meson loop contributions to the  $D^* D \pi$ form factor: 
``self energies''.}
\label{fig3}
\end{figure}

\begin{figure} [h]
\centerline{\epsfig{figure=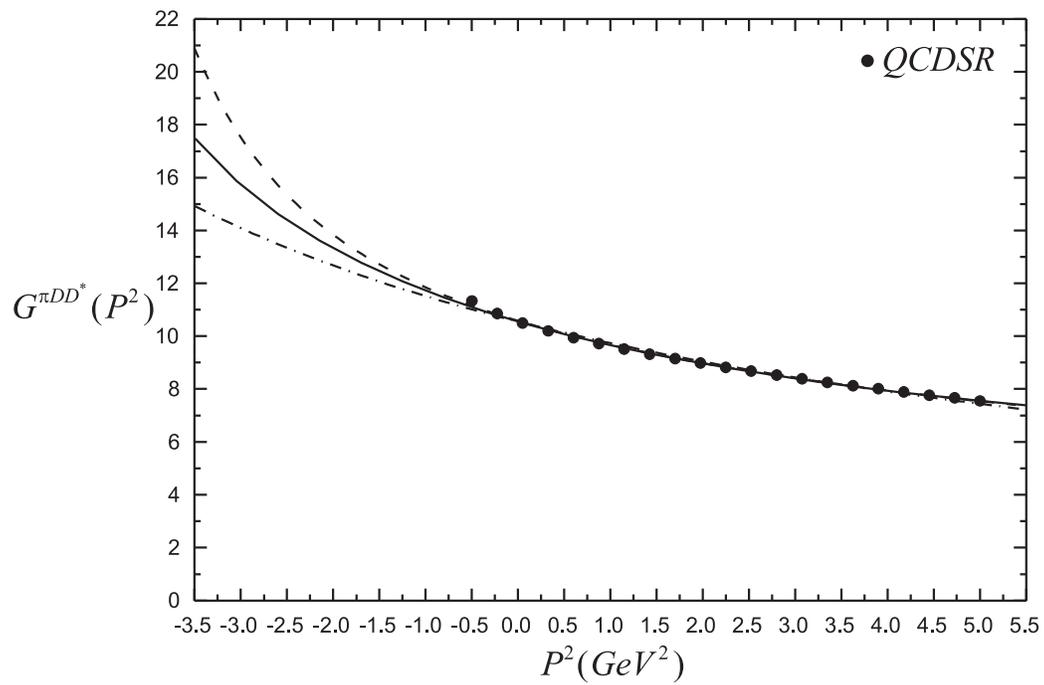,width=14.cm}}
\caption{The $D^* D \pi$ form factor. Dots: QCDSR from \protect\cite{revisit}; 
solid, dash and dash-dotted lines are fits obtained with eq. (\protect\ref{gfinal}),  (\protect\ref{GFI}) and (\protect\ref{GFII}), respectively.}
\label{fig5}
\end{figure}

\begin{figure} [h]
\centerline{\epsfig{figure=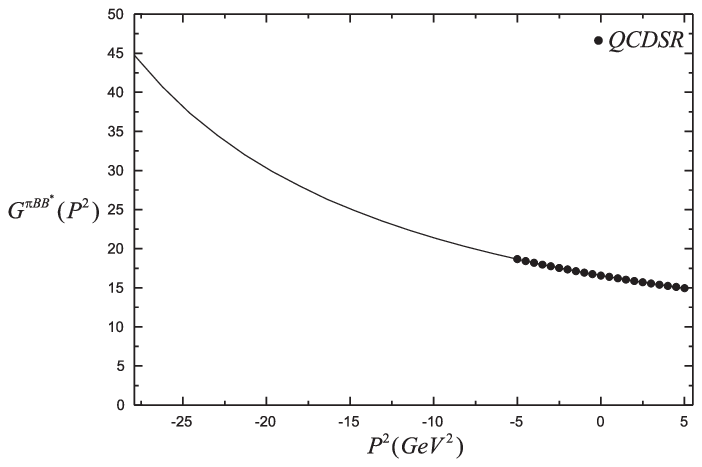,width=14.cm}}
\caption{The same as Fig. 5 for the $B^* B \pi$ form factor.}
\label{fig6}
\end{figure}


\begin{thebibliography}{99}


\bibitem{cleo} S. Ahmed {\it et al.}, [CLEO Collaboration],
{\sl Phys. Rev. Lett.} {\bf 87}, 251801 (2001)
{\tt [hep-ex/0108013]}; 
A. Anastassov {\it et al.}, [CLEO Collaboration],
{\sl Phys. Rev.} {\bf D65}, 032003 (2002)
{\tt [hep-ex/0108043]}.


\bibitem{col} P. Colangelo {\it et al.}, {\sl Phys. Lett.} {\bf B339}, 151 (1994).

\bibitem{ele} V.L. Eletsky and Ya.I. Kogan, {\sl Z. Phys.}  {\bf C28}, 155 
(1985);
A.A. Ovchinnikov, {\sl Sov. J. Nucl. Phys.} {\bf50}, 519 (1989).


\bibitem{bel} V.M. Belyaev {\it et al.}, {\sl Phys. Rev.} {\bf D51}, 6177 (1995). 

\bibitem{col2} P. Colangelo and F. De Fazio, {\sl Eur. Phys. J.} {\bf C4},
503 (1998).

\bibitem{kho} A. Khodjamirian {\it et al.}, {\sl Phys. Lett.} {\bf B457}, 245 
(1999).

\bibitem{gro} A.G. Grozin and O.I. Yakovlev, {\sl Eur. Phys. J.} {\bf C2},
721 (1998).

\bibitem{dn} H.G. Dosch and S. Narison, {\sl Phys. Lett.} {\bf B368}, 163 
(1996).

\bibitem{ddpi}  F.S. Navarra, M. Nielsen, M.E. Bracco, 
                M. Chiapparini and C.L. Schat, 
                , {\sl Phys. Lett.} {\bf B489}, 319 (2000). 

\bibitem{dirac} D. Becirevic and A. LeYaouanc, {\sl JHEP} {\bf 9903}, 021 (1999).

\bibitem{herdoiza} 
A. Abada, D. Becirevic, P. Boucaud, G. Herdoiza, J.P. Leroy, A. LeYaouanc, O. P\`ene, 
J. Rodriguez-Quintero, {\sl Phys. Rev.} {\bf D66}, 074504 (2002).

\bibitem{damir} D. Becirevic, J. Charles, A. LeYaouanc, L. Oliver, O. P\`ene and 
                J.C. Raynal, {\tt [hep-ph/0212177]}.


\bibitem{kimlee} H.C. Kim, Su Houng Lee, {\sl Eur. Phys. J.} {\bf C22}, 707 (2002).


\bibitem{kim} H.C. Kim, {\sl J. Korean  Phys. Soc.} {\bf 42}, 475 (2003);
              hep-ph/0206170. 


\bibitem{ddrho} M.E. Bracco, M. Chiapparini, A. Lozea, F.S. Navarra and M. Nielsen, 
                {\sl Phys. Lett.} {\bf B521}, 1 (2001).

\bibitem{revisit} F.S. Navarra, M. Nielsen and M.E. Bracco, {\sl Phys. Rev.} 
                  {\bf D65}, 037502 (2002).


\bibitem{speth} A. Szczurek, H. Holtmann and J. Speth, {\sl Nucl. Phys.} {\bf A605}, 
                496 (1996).


\bibitem{linko} Z. Lin and C.M. Ko,  {\sl Phys. Rev.} {\bf  C62}, 034903  (2000). 


\bibitem{mamu98} S.G. Matinyan and B. M\"uller, {\sl Phys. Rev.} {\bf C58}, 2994  
                 (1998).

\bibitem{osl} Y. Oh, T. Song and S.H. Lee, {\sl Phys. Rev.} {\bf C63},  034901 (2001). 
\bibitem{mane1}  F.S. Navarra, Marina Nielsen and M.R. Robilotta, 
                 {\sl Phys. Rev.} {\bf C64}, 021901 (2001). 

\end{thebibliography}
\end{document}